\documentclass[twocolumn]{revtex4}

\usepackage{graphicx}
\usepackage[colorlinks=true,citecolor=black,urlcolor=black]{hyperref}

\begin{document}

\title{Photoluminescence of patterned arrays of vertically stacked InAs/GaAs quantum dots}

\author{T. W. Saucer}
\affiliation{Department of Physics, University of Michigan, Ann Arbor, MI 48109}
\author{J.-E. Lee}
\affiliation{Department of Physics, University of Michigan, Ann Arbor, MI 48109}
\author{A. J. Martin}
\affiliation{Department of Materials Science and Engineering, University of Michigan, Ann Arbor, MI 48109}
\author{D. Tien}
\affiliation{Department of Physics, University of Michigan, Ann Arbor, MI 48109}
\author{J. M. Millunchick}
\affiliation{Department of Materials Science and Engineering, University of Michigan, Ann Arbor, MI 48109}
\author{V. Sih}
\affiliation{Department of Physics, University of Michigan, Ann Arbor, MI 48109}

\begin{abstract}
We report on photoluminescence measurements of vertically stacked InAs/GaAs quantum dots grown by molecular-beam epitaxy on focused ion beam patterned hole arrays with varying array spacing.  Quantum dot emission at 1.24 eV was observed only on patterned regions, demonstrating preferential nucleation of optically-active dots at desired locations and below the critical thickness for dot formation at these growth conditions.  Photoluminescence measurements as a function of varying focused ion beam irradiated hole spacing showed that the quantum dot emission intensity increased with decreasing array periodicity, consistent with increasing dot density. 
\end{abstract}

\maketitle

\section{Introduction}
Semiconductor quantum dots (QDs) are of interest for applications ranging from solar cells and optical devices to quantum computing.  InAs QDs are typically grown by self-assembly, which leads to nucleation at random spatial locations and a range of dot sizes.  However, the ability to create localized patterns of dots could lead to improved device functionality.  For example, optical devices could be developed with spatially varying optical responses for multi-wavelength sources and detectors.  Additionally, some quantum computing proposals rely upon coupling QDs to photonic crystal cavities, which requires precise dot positioning to achieve good spatial and spectral overlap between the QD and cavity mode \cite{englund05, badolato05}.  Recent work, has shown the ability to create patterned arrays of self-assembled QDs by both ex situ lithography and etching \cite{schneider08, atkinson08} and {\it in vacuo} patterning \cite{mckay07a, mckay07b, mehta07, lee09}. In contrast to {\it in vacuo} techniques, lithographic methods require cleaning to remove contaminants and oxidation between patterning and sample growth.   In general, QD formation may be induced in focused ion beam (FIB) irradiated regions despite the fact that the deposited thickness is less than the critical thickness for their formation under typical growth conditions. Furthermore, parameters such as dot height and diameter are largely unaffected by the ion beam dose \cite{lee09}.   By eliminating QD formation outside of the patterned region, FIB-directed QD growth has the potential to deterministically place optically active QDs on chip with desired QD spacing.  In this paper, we report photoluminescence measurements of patterned, multilayer QD structures as a function of FIB hole array spacing (henceforth referred to as ``array spacing''), FIB ion dose, and temperature.
\section{Growth parameters and experimental methods}
\begin{figure}
\includegraphics[width=\linewidth]{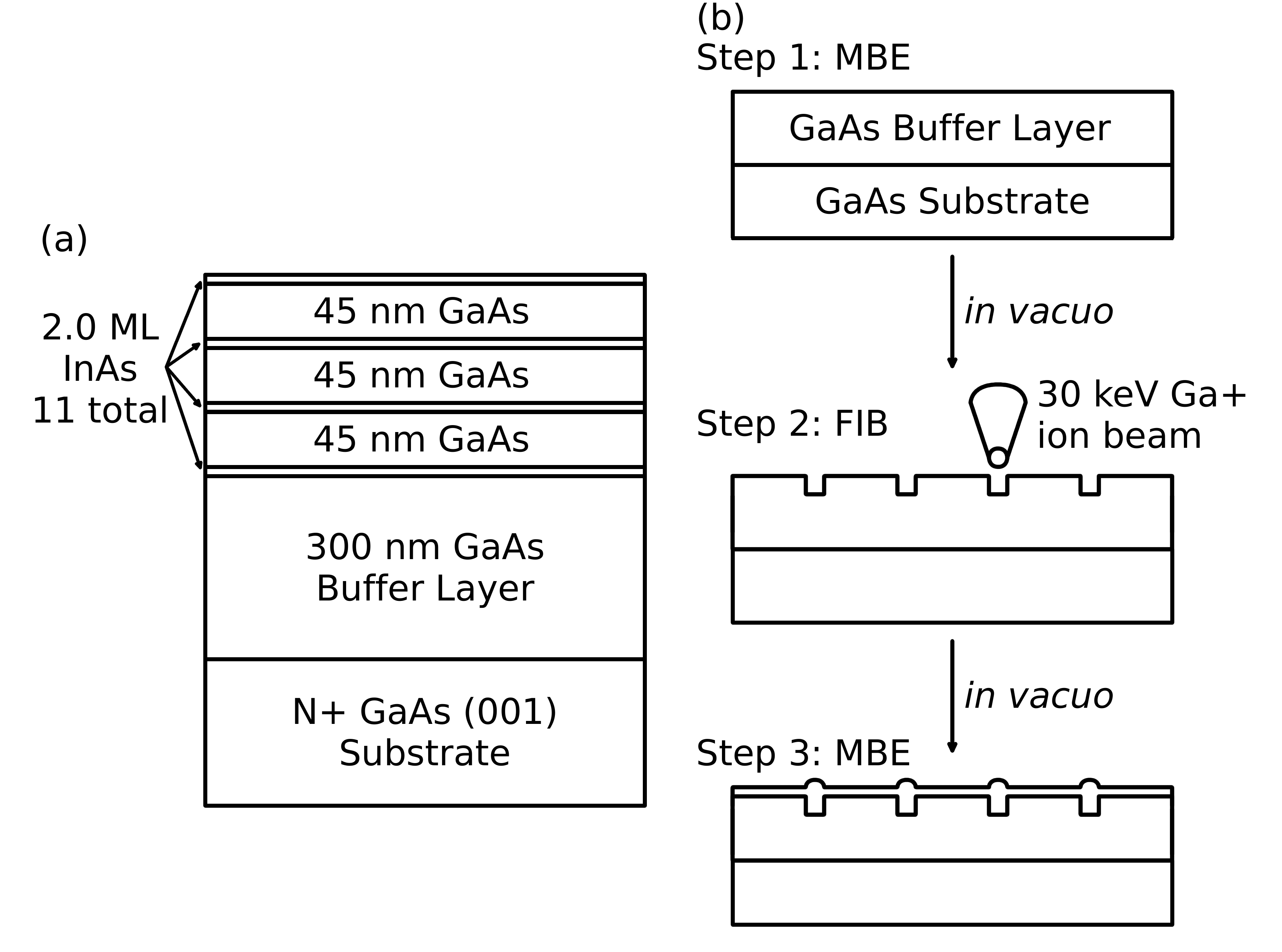}
\caption{(a) Schematic of the sample layer structure.  (b) Sample fabrication.  In step 1, a 300 nm GaAs buffer layer is grown and then annealed for 10 minutes.  In step 2, the sample is transferred {\it in vacuo} to the FIB chamber for patterning.  In step 3, the sample is transferred {\it in vacuo} back to the MBE for layer growth.}
\end{figure}
	The QD multilayer sample was grown by molecular beam epitaxy (MBE) on an N+ GaAs(001) substrate that was patterned using an {\it in vacuo} FIB.  A schematic of the sample structure and fabrication steps is shown in Figure 1.  Following oxide desorption, a 300 nm GaAs buffer layer was grown at a substrate temperature of T = 590$^\circ$C. All substrate temperatures were determined using an optical pyrometer. The As and Ga fluxes for buffer growth were 2.8 and 1.0 monolayers s$^{-1}$ (ML s$^{-1}$) respectively, as determined by reflection high energy electron diffraction. Following buffer growth, the sample was annealed for 10 minutes and then quenched under As overpressure. The sample was subsequently transferred {\t in vacuo} to the FIB for patterning. The patterns consisted of circular, FIB-milled holes in a $40 \times 40 \mu\text{m}^2$ square array, separated by four different spacings of 0.25, 0.5, 1.0, and 2.0 $\mu$m, each at dwell times of both 1.0 and 3.0 ms, for a total of 8 different patterns.  Based on our previous work \cite{lee09} and AFM measurements of samples prepared under similar conditions, we estimate the FIB holes are approximately 40 nm in diameter and 2-5 nm deep.  Each hole was dosed in a single pass with a 9.2 pA, 30 keV Ga+ ion beam. After patterning, the sample was transferred back to the MBE chamber and the sample temperature was raised to the growth temperature of T = 485$^\circ$C under an As flux of 2.7 ML s$^{-1}$.  Eleven layers of 2 ML thick InAs were deposited at a growth rate of 0.11 ML s$^{-1}$, with a 45 nm thick GaAs interlayer grown at a deposition rate of 1.0 ML s$^{-1}$.  After growth of the 11th layer of InAs, the sample was immediately quenched under an As overpressure.  The templating and growth conditions were selected based on our previous work which has shown that the holes direct the preferential nucleation of quantum dots \cite{lee09}.

For photoluminescence (PL) measurements, the samples were mounted in a helium flow cryostat and optically excited with 7.5 $\mu$W incident power from a HeNe laser operating at 633 nm.  The excitation beam was focused to a 30 $\mu$m diameter spot, resulting in an intensity of $10^4$ W/m$^2$.  We used a confocal microscope configuration, which allowed us to image the sample and collect from only the FIB-patterned region of interest.  The aperture size was adjusted such that the collection region is approximately 20 $\mu$m in diameter, which is smaller than each FIB-patterned region ($40 \times 40 \mu\text{m}^2$).  We measured the PL spectra using a 0.75 m spectrometer and a liquid nitrogen cooled InGaAs detector.
\section{Results and discussion}
\begin{figure}
\includegraphics[width=\linewidth]{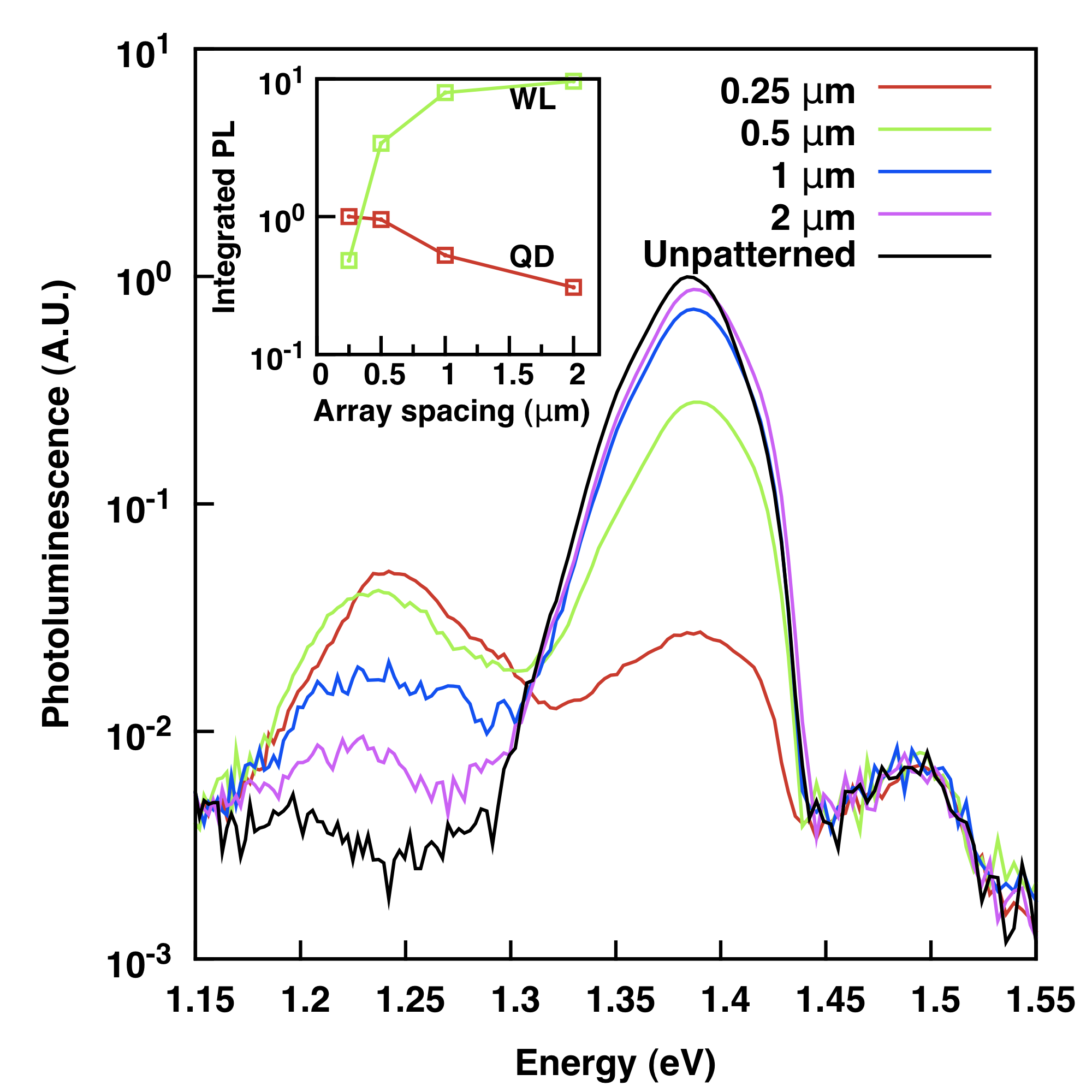}
\caption{Photoluminescence spectra for varying FIB hole array spacing (1.0 ms dwell time) at 30 K.  Inset: Integrated PL from the quantum dot and wetting layer peaks as a function of array spacing.}
\end{figure}
\begin{figure}
\includegraphics[width=\linewidth]{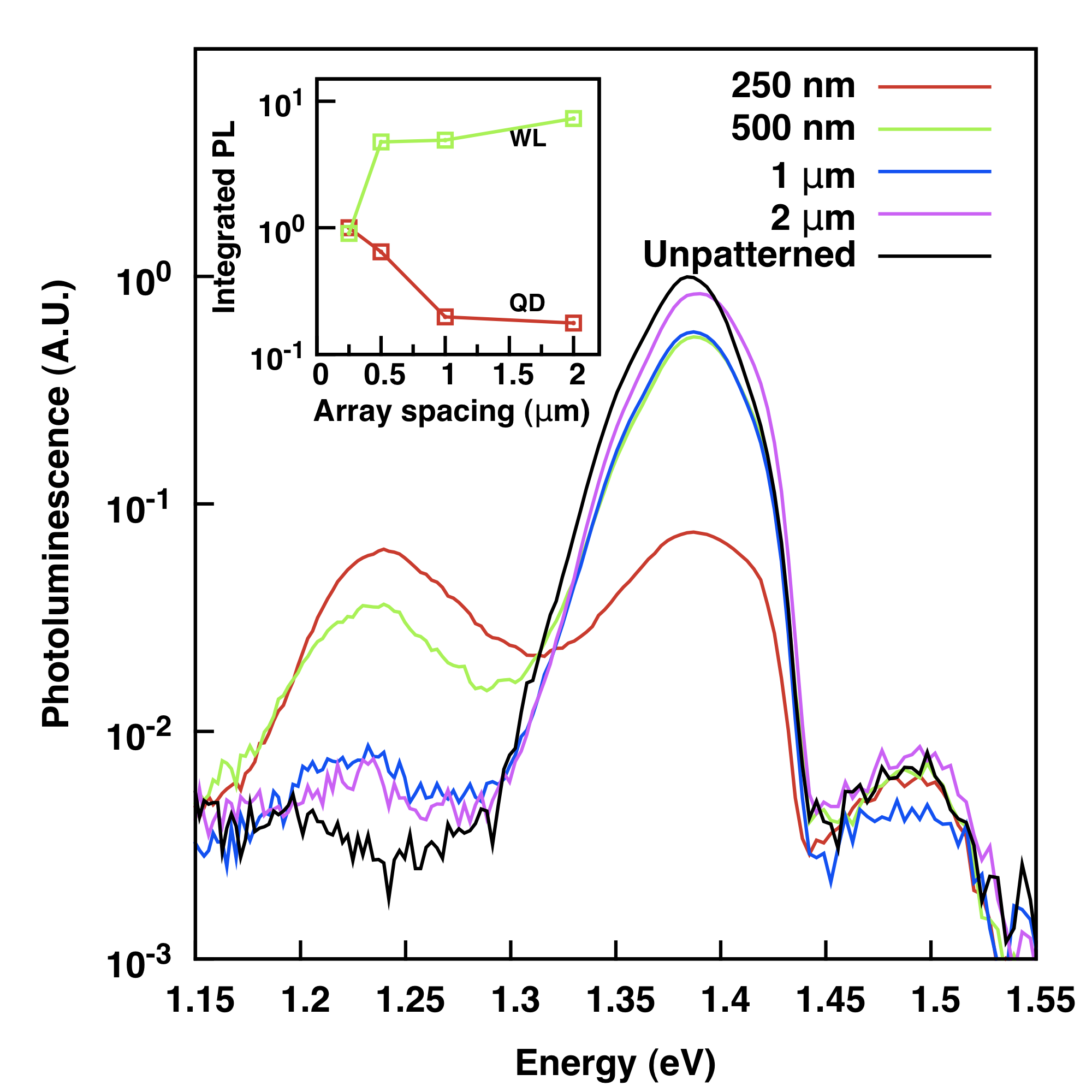}
\caption{Photoluminescence spectra for varying FIB hole array spacing (3.0 ms dwell time) at 30 K.  Inset: Integrated PL from the quantum dot and wetting layer peaks as a function of array spacing.}
\end{figure}
	PL measured from all eight patterned regions and an unpatterned area of the sample is shown in Figure 2 (1.0 ms dwell time) and Figure 3 (3.0 ms), which are normalized to the same unpatterned region.  The data was taken at a temperature T = 30 K and shows three distinct peaks centered around 1.24, 1.38, and 1.5 eV, corresponding to the InAs QDs, InAs wetting layer, and GaAs substrate, respectively.  We observe that the intensity of QD emission at 1.24 eV decreases monotonically with increasing array spacing, which is consistent with decreasing QD density.  In the unpatterned region, we do not observe a peak at 1.24 eV, confirming that QDs only form within the patterned regions despite the fact that the critical thickness has not been exceeded.  In addition, the QD emission peak does not shift significantly with changing array spacing, indicating that the size of the QDs are not greatly affected by that parameter.  The wetting layer peak decreases in intensity as the array spacing decreases, as expected assuming material from the wetting layer is depleted by the creation of QDs.  In addition, photoexcited carriers in the wetting layer may be captured by the QDs, which would also lead to the observed decrease in emission from the wetting layer as QD density increases.  The inset shows the integrated QD PL as a function of FIB hole spacing.  Considering only the effect of QD density, as the spacing decreases by half, the dot density should increase by a factor of four leading us to expect a four-fold increase in emission.  However, decreasing the array spacing from 2 $\mu$m to 1 $\mu$m and from 1 $\mu$m to 0.5 $\mu$m only increases the emission by a factor of about 1.8, and decreasing array spacing from 0.5 $\mu$m to 0.25 $\mu$m results in only a minor increase in emission.  Supposing that the diffusion length of photoexcited carriers is larger than the array spacing, we expect that further decreases in array spacing should not have much effect on the PL intensity.

	To determine how the FIB templating conditions affect QD PL, we compare the data in Figure 2 and Figure 3, measured in regions with 1.0 ms and 3.0 ms dwell times, respectively. In both cases, QD emission decreases and wetting layer PL increases with increasing array spacing.  While the QD intensity in the regions with 0.25 $\mu$m and 0.5 $\mu$m array spacing is similar for both 1.0 ms and 3.0 ms dwell times, the QD emission is significantly lower for the 1 $\mu$m and 2 $\mu$m array spacings for the longer dwell time.  We hypothesize that this discrepancy could arise from the deeper holes or increased ion beam damage created by the increased dwell time \cite{lee09} that make it more difficult to form optically active QDs for the larger array spacings.  
\begin{figure}
\includegraphics[width=\linewidth]{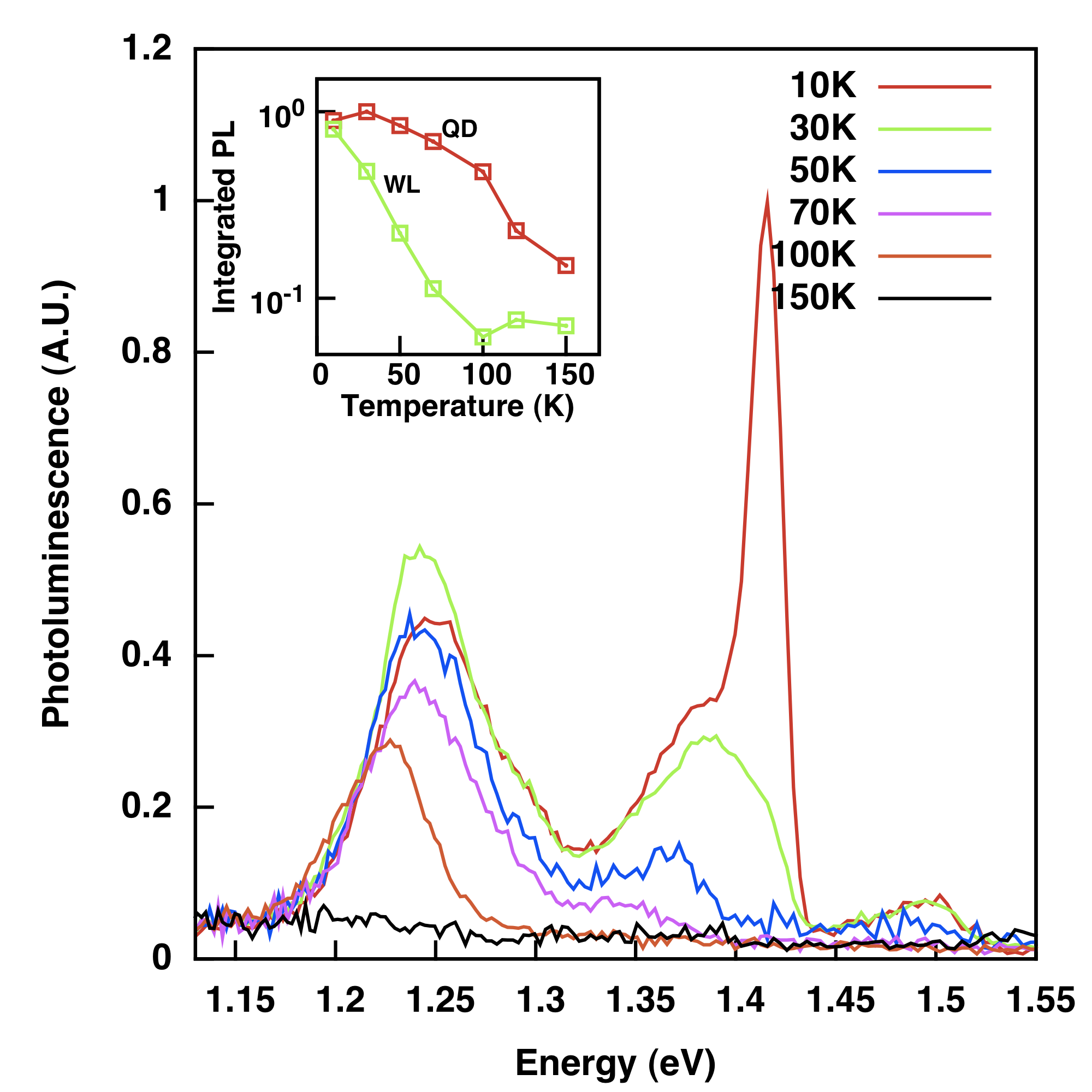}
\caption{Quantum dot photoluminescence on the 0.25 $\mu$m spaced patterned array for temperatures 10-150 K.  Inset:  Integrated PL intensity from QD and wetting layer peaks as a function of temperature.}
\end{figure}

The full width half maximum of the QD peak is approximately 80 meV, as determined from fitting of the data from the 0.25 $\mu$m and 0.5 $\mu$m array spacing patterns.  Since the linewidth is an indication of the inhomogeneity of the QDs, it is a useful measure of the effect of the FIB templating on the QD size distribution.  Without optimizing growth our linewidth is larger than the typical inhomogeneous linewidth of 50 meV reported for unpatterned self-assembled InAs QDs \cite{dai97, chu99, brusaferri96, grundmann95}.  Our previous atomic force microscopy characterization of similar uncapped FIB-templated QDs showed QD diameters of $60 \pm 30$ nm and heights of $16 \pm 8$ nm \cite{lee09}.  However, we expect that this inhomogeneity can be reduced through further optimization of the templating, growth conditions, or post growth rapid thermal annealing \cite{malik97, xu98}.

Temperature dependent PL provides further confirmation that the peak at 1.24 eV is QD emission and that the peak at 1.38 eV is from the wetting layer.  Figure 4 shows PL from the region with 0.25 ?m array spacing and 1 ms dwell time from 10-150 K.  The inset shows the integrated intensity of the QD and wetting layer peaks as a function of temperature.  We find that the QD PL is at a maximum at 30 K and that the wetting layer PL decreases more rapidly with increasing temperature.  This decrease in intensity at high temperatures is expected as non-radiative recombination, or thermal quenching, becomes increasingly more significant.  This thermal quenching occurs more rapidly in the wetting layer, since QDs have delta like density of states \cite{leru03}.  The QD emission energy decreases with increasing temperature, as expected.
\section{Conclusion}
	We have demonstrated spatial control of the optical response of a material by deterministic placement of vertically stacked InAs/GaAs QDs using FIB templating and MBE growth.  We find that changing the array spacing significantly affects the magnitude of QD and wetting layer emission.  We measured the PL as a function of temperature, finding maximum QD emission at 30 K.
\section{Acknowledgement}
This work was supported by the University of Michigan.  JMM and AJM acknowledge the support of the EFRC grant number DE-SC000957.

\end{document}